\documentclass[aps,showpacs,twocolumn,floatfix]{revtex4} 
\usepackage{graphicx}

\begin{document}
   
 
\title{Simulation of a stationary dark soliton in a trapped
zero-temperature Bose-Einstein
condensate}
\author{Sadhan K. Adhikari\thanks{e-mail: adhikari@ift.unesp.br}}
\affiliation{Instituto de F\'{\i}sica Te\'orica, UNESP - S\~ao Paulo
State University, 01.405-900 S\~ao Paulo, S\~ao Paulo, Brazil}

\date{\today}

\begin{abstract}

We discuss a computational mechanism for the generation of a stationary
dark soliton, or black soliton, in a trapped Bose-Einstein condensate
using the Gross-Pitaevskii (GP) equation for both attractive and repulsive
interaction.  It is demonstrated that the black soliton with a ``notch" in
the probability density with a zero at the minimum is a stationary
eigenstate of the GP equation and can be efficiently generated numerically
as a nonlinear continuation of the first vibrational excitation of the GP
equation in both attractive and repulsive cases in one and three
dimensions for pure harmonic as well as harmonic plus optical-lattice
traps.  We also demonstrate the stability of this scheme under different
perturbing forces.

\end{abstract}
\pacs{03.75.Lm, 05.45.Yv}
\maketitle

\section{Introduction} Solitons are solutions of wave equation where
localization is obtained due
to a nonlinear interaction and  have been observed in optics
\cite{1},  water waves \cite{1}, and in
Bose-Einstein
condensates (BEC) \cite{2,vr,3}. The bright solitons of BEC represent
local
maxima \cite{4} and
the dark and grey solitons local minima \cite{bur,5a,5b,5c}. A 
stationary dark soliton 
where the local minimum goes to zero value is called a
black soliton. 
There
have
been
experimental
study  of bright \cite{3},  dark and grey \cite{2,vr}
solitons
of
BEC. Dark solitons of
nonlinear optics \cite{1} are governed by the
nonlinear
Schr\"odinger (NLS) equation  which is  similar to the mean-field
Gross-Pitaevskii (GP) equation \cite{7}  describing a trapped BEC. More
recently, dark solitons have been observed in trapped BECs \cite{2,vr}.

The one-dimensional NLS equation in the repulsive
or
self-defocusing case is usually written as \cite{1}
\begin{equation}\label{nls} 
i u_t+u_{xx}-  |u|^2u=0,
\end{equation}
where the  time
($t$) and space ($x$) dependences of the wave function $u(x,t)$ are
suppressed.
This equation sustains the following
 dark and grey solitons \cite{5}:
\begin{eqnarray}\label{DS}
u(x,t)=r(x-ct) \exp[-i\{\phi(x-ct)-\mu t   \}],
\end{eqnarray}
with
\begin{eqnarray}
r^2(x-ct)& = & \eta -2\xi^2 \mbox{sech}^2[\xi(x-ct)],  \\
\phi(x-ct)&=&\tan^{-1}[-2 \xi/c \hskip 0.05cm \tanh\{\xi (x-ct)\}],
\\
 \xi &=& \sqrt{(2\eta - c^2)}/2, 
\end{eqnarray}
where $c$ is the velocity, $\mu$ is a parameter,
and $\eta$ is related to intensity. Soliton (\ref{DS})
having  a ``notch" over a
background density is grey in general. It is dark if density
$|u|^2=0$ at the minimum.
At zero velocity the soliton becomes a dark soliton:
$|u(x,t)|= \sqrt {\eta} \tanh [\sqrt{(\eta/2)}x]$.

The similarity of the NLS equation (\ref{nls}) to the GP equation of a
trapped BEC (Eq. (\ref{d1}) below) imply the possibility
of a
stationary dark soliton, or a black soliton, 
in a trapped zero-temperature BEC \cite{bur}.  It has been suggested that
the black 
soliton of a trapped BEC could be a stationary eigenstate of the GP
equation \cite{bur,5b} as in the case of the trap-less NLS equation.  
Here we re-investigate the origin of the black  soliton
in
a trapped  zero-temperature BEC and 
 point out that this soliton 
\cite{5a,5b,8,9,10} is 
the first  vibrational  excitation of the GP equation for both attractive
and repulsive atomic interactions
and is a
stationary eigenstate. 
We suggest a scheme for numerical simulation 
of a stationary dark soliton by time evolution of the linear  GP equation
starting with the  analytic vibrational excitation, 
while the nonlinearity
is slowly introduced. 
We simulate a  stationary dark soliton in  a harmonic and harmonic plus
optical-lattice traps in one and three dimensions. 
In all cases  the simulation 
proceeds through successive eigenstates 
of the GP equation. 
Consequently, the  stationary dark soliton in a trapped BEC could be  kept
stable 
during  numerical simulation. 
To illustrate the stability of our scheme we also study the breathing
oscillation of the stationary dark soliton upon application of
different perturbations.

The  stationary dark soliton being an excited state is thermodynamically 
unstable. It is also unstable due to quantum fluctuations
\cite{z}. However, these instabilities are not manifested in the
mean-field
model.  
It seems that it will be difficult to generate black  solitons 
experimentally
because they are fragile to perturbations, transverse dimensions, quantum
effects, and thermal perturbations, etc.  Nevertheless, we demonstrate
that they are stationary eigenstates of the GP equation and exploit this
information to illustrate a simple numerical scheme for their generation. 
The present numerical scheme has recently been used successfully
to simulate
the 
 stationary dark  solitons of a degenerate boson-fermion mixture
\cite{bf}.

\section{Black  Soliton in a Harmonic Trap} The mean-field dynamics of a
trapped
BEC
is
usually described by the time-dependent GP 
equation \cite{7}. 
For  a strong radial confinement in an axially-symmetric
configuration, the GP equation can be reduced to the following 
quasi-one-dimensional  form \cite{8,9,10}
\begin{eqnarray}\label{d1} 
i u_t+u_{xx}- n |u|^2u=V(x)u, 
\end{eqnarray}
where  a positive  nonlinearity $n$
represents repulsive (self-defocusing) interaction
and a negative $n$ represents 
attractive
(self-focusing)
interaction. 
In  Eq. (\ref{d1}) $V(x)$ is the external 
trapping potential. 
The normalization of the
wave function is given by
$\int_{-\infty}^{\infty}|u|^2 dx =1$.
The reduction of the GP equation from three to one dimension 
can be performed in a straightforward fashion for a single- \cite{abdul}
as
well as coupled-channel \cite{abdul1} cases for small
nonlinearity. Nevertheless, for large nonlinearity corrections are needed
\cite{abdul2}. However, we shall neglect these corrections here.

There is no known analytic
solution to  Eq. (\ref{d1}) for $V(x)\ne 0 $ and $n\ne 0$. Of course, for
$n=0$ we have the well-known harmonic oscillator solution for a harmonic
trap.  However, for  $V(x)=0$ and
positive $n$,   Eq. (\ref{d1}) has the following unnormalizable dark
soliton:   
\begin{equation}\label{ds1}
u(x,t)= \sqrt{2/n} \tanh (x) \exp{(-2it)}.
\end{equation}
Soliton  (\ref{ds1}) has a stationary
notch with zero  minimum at $x=0$ on a constant
background  extending to $x=\pm \infty$.

It has been conjectured that a stationary normalizable 
dark soliton 
exists in Eq. (\ref{d1})  with a  harmonic trap:
$V(x)=x^2$
and satisfies the same boundary condition as 
 \cite{5c,8,9,10}
\begin{equation}\label{ds2}
u_{\mbox{\small DS}}(x)= N  \tanh (x) u_S(x) ,
\end{equation}  
where $u_{\mbox{\small DS}}(x)$ is the black  soliton, $ u_S(x)$  the
ground-state 
solution   to Eq. (\ref{d1}) and $N$ the normalization. 
The form (\ref{ds2}) has been used \cite{9} as an
initial guess
to a fixed point algorithm that finds the exact numerical
stationary solution.

The ansatz (\ref{ds2}) has been used as starting guess 
in  numerical studies on  dark
solitons. Actually, in some applications \cite{9}
the  Thomas-Fermi (TF) approximation
$u_{\mbox{\small TF}}$ \cite{7} has been used in place of $u_S$ 
for positive $n$:
\begin{equation}\label{TF}
u_S(x) \approx u_{\mbox{\small TF}}(x)=  \sqrt{\mbox{max}(0,[\mu-V(x)]/n
)  },
\end{equation}
where  max(,) denotes the larger of the
two
arguments
and 
$\mu$ is the chemical potential for  solution $ u_S(x)$.

The ansatz   (\ref{ds2})  is not an eigenstate of 
Eq.  (\ref{d1}).  Assuming that it is close  to an 
eigenstate,  numerical iteration of  Eq. (\ref{d1})     
should lead to  a  stationary  dark soliton  at large times. 
Whenever the input  
(\ref{ds2}) is not a good  approximation to the
  stationary dark soliton, oscillations are expected upon iteration. 
Kevrekidis {\it et al.} \cite{9} give 
a detailed
parametric linear stability analysis of the stationary solution
obtained through their fixed point iteration scheme varying 
different parameters, and find 
windows
of stability and windows of instability. 
They also extend their discussion
on instability to the case of an optical-lattice potential.
In many  cases no  stable soliton
has
been obtained, or  an oscillating grey soliton has been found
\cite{9}. 

One way to achieve a stable soliton with initial guess (\ref{ds2}), which
is not an eigenstate of Eq. (\ref{d1}), is to include a dissipation in the
system.  The numerical iteration of a slightly inaccurate solution would
generate radiation in general and  usually not converge to any stationary
state without dissipation. This could be the source of instability in
Ref. \cite{9}. In the
following, first we study the deficiency of using Eq. (\ref{ds2}) in
Eq. (\ref{d1})  without
dissipation in generating a   stationary  dark soliton and then
illustrate the
present alternative scheme, where we use an exact eigenstate of
Eq. (\ref{d1}) to generate the   stationary dark soliton.

First  we performed extensive  
calculations using ansatz
(\ref{ds2})  in the time evolution of Eq.  (\ref{d1}) using the
Crank-Nicholson algorithm  \cite{11a} for
different $n$.
We discretize the NLS
equation with space step 0.05 
and time step 0.0025, which was enough for
achieving convergence of a stationary problem.
We used 
accurate
numerical solution to $u_S(x)$ in place of the TF  approximation
(\ref{TF}). The present Crank-Nicholson algorithm is appropriate
not only for the calculation of stationary states but also
for nonequilibrium dynamics \cite{11a} with absorptive potential during
collapse  \cite{collapse}.

\begin{figure}
 
\begin{center}
\includegraphics[width=\linewidth]{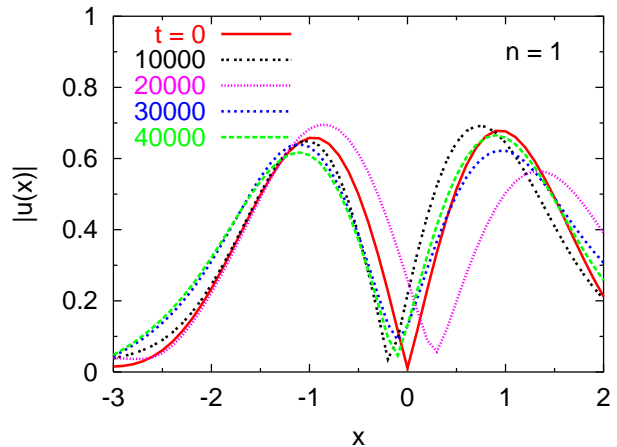}
\end{center}

\caption{The   dark soliton   $|u(x)|$ 
of Eq. 
(\ref{d1}) with  $V(x)=x^2$
vs. 
$x$  for   $n=$   1  obtained by
time evolution with  input (\ref{ds2}) at $t=0$, 10000,
20000, 30000, and  40000. The dark soliton oscillates as a grey 
soliton without ever converging.
}
\end{figure}

\begin{figure}[!ht]
 
\begin{center}
\includegraphics[width=.47\linewidth]{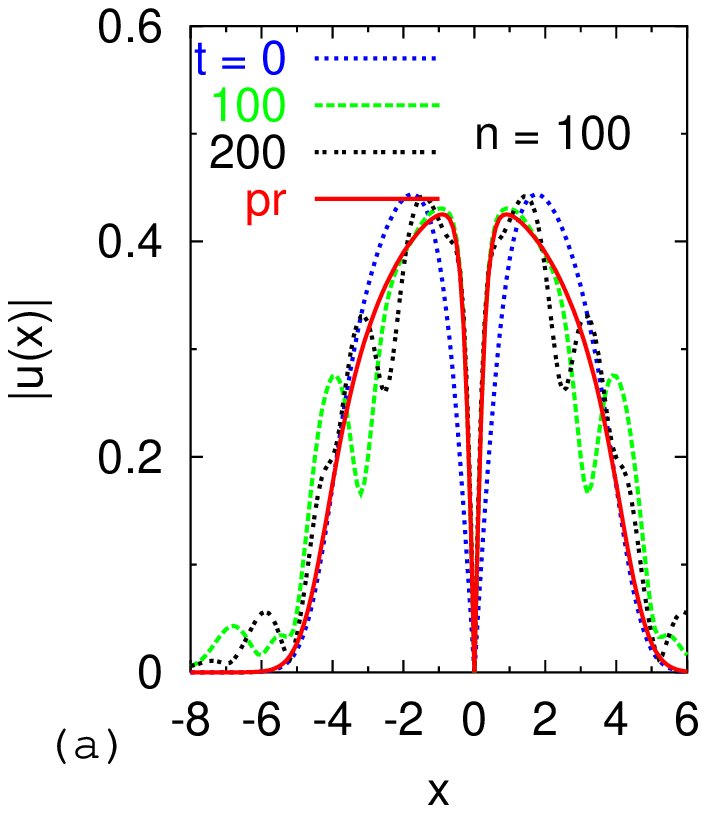}
\includegraphics[width=.47\linewidth]{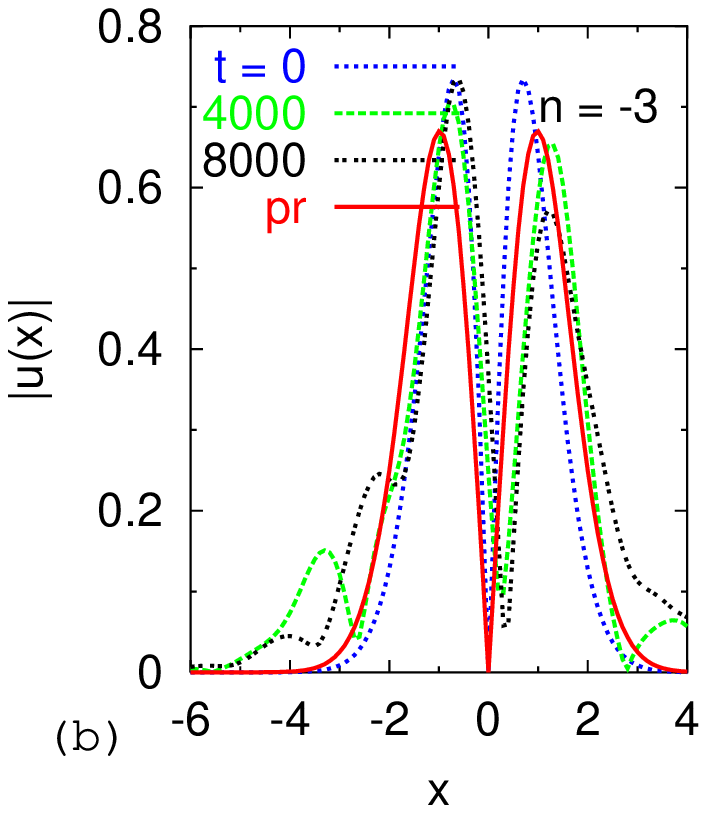}
\end{center}
 
\caption{The dark soliton    $|u(x)|$ 
of Eq.
(\ref{d1}) with  $V(x)=x^2$
vs. 
$x$  for $n=$ (a)  100 and (b) 
$-3$  obtained by
time evolution with  input (\ref{ds2}) at  different times $t$.
Also plotted are  present stationary  results
obtained with input (\ref{ex}) marked ``pr"
(full red  line) which do not change with time.
}

\end{figure}

In place of the Crank-Nicholson numerical scheme used in this paper one
could also use the method of imaginary time propagation \cite{itime} for
obtaining the
stationary states of the GP equation. By replacing the time by an
imaginary time variable, the original time-dependent GP equation becomes a
diffusion-like equation, and propagation in imaginary time leads to
relaxation towards the vibrational ground state. The imaginary time
propagator can be expanded in Chebychev polynomials which leads to a
stable and efficient scheme. This approach can be easily modified to
obtain vibrationally excited states as well. By repeating the relaxation
(imaginary time propagation) but filtering out any contribution of the
ground state at each time step with anti-periodic boundary conditions, one
obtains the first vibrationally excited state. This could be an
interesting future work.

Now  we perform  a direct time evolution of the full GP equation with
ansatz (\ref{ds2}) as input  and show in Fig.  1 the solution at
different times for $n=1$. The authors of Ref. \cite{9} are making a
different time evolution, e. g.,  a perturbation of their exact stationary  
solitary wave with a uniformly distributed random field of
amplitude 0.01. As the successive states are not stationary eigenstates 
these schemes may run into numerical difficulty when the nonlinearity is
large. Even for a relatively small nonlinearity of $n=1$,
in that approach
the solution does not converge at large times: the initial black
soliton becomes a grey soliton and oscillates around a mean position at
the center of the trap.
Although the wave-function density is symmetric at $t=0$, it becomes
non-symmetric with the evolution of time. This makes the 
black  soliton to
oscillate as a grey soliton  upon time evolution before being destroyed 
eventually for much larger values of time $t$. 
This trouble as noted in Fig.  1 increases
with the increase of nonlinearity $n$.

To circumvent  the above-mentioned problem, 
we  find a direct solution to
Eq. (\ref{d1}) for the   stationary dark soliton with the asymptotic
boundary condition
implicit in Eq. (\ref{ds2}), e. g.,  
$u_{\mbox{\small DS}}(x) \sim x$ as $x\to 0$ and
$u_{\mbox{\small DS}}(x) \to
0 $
as $x
\to \pm \infty$. The   solution satisfying these
conditions is the
nonlinear evolution of the first vibrational excitation  of the linear
oscillator, 
obtained by setting  $n=0$  in  Eq. (\ref{d1}):
\begin{equation}\label{ex}
u_1(x,t)=\sqrt{(2/\sqrt \pi)} x \exp(-x^2/2)\exp(-3it). 
\end{equation}
The possible   stationary 
dark soliton of  Eq. (\ref{d1}) can be obtained
by time
evolution  of the GP equation 
with $u_1(x,0)$ as input at $t=0$, setting
$n=0$. During time evolution the  nonlinearity
$n$  should be slowly introduced until the desired nonlinearity  
is achieved. In this work we increased $n$ by 0.001 at each time
propagation, which was sufficient for convergence.
By this procedure a stationary  dark soliton
could be  obtained for very large $n$.

Next we compare the time evolution of the dark soliton using conventional
ansatz (\ref{ds2}) and the present suggestion based on Eq. (\ref{ex}). The
results of numerical simulation using the two schemes are plotted in Figs.
2 (a) and (b) for $n=100$ and $-3$ at different times.  We show results
for these two $n$, as we found that convergence was more difficult for
negative $n$ and large positive $n$ values. The case $n=-3$ discussed here
is specially interesting as we demonstrate, contrary to popular belief,
that dark solitons of a trapped quasi-one-dimensional BEC could be
stationary for attractive interactions as well.  The iterative solution
using Eq. (\ref{ds2})  may execute oscillation on time evolution, whereas
the solution from input (\ref{ex}) remains stationary on time evolution as
the system passes through successive eigenstates and results in a
stationary dark soliton of the full NLS equation. In Figs.  2 we show the
result for the soliton using the present procedure only at $t=0$ as this
result does not change with time. The result based on Eq. (\ref{ds2})
oscillates on time evolution as can be found from Figs. 2. However, the
oscillation is not so severe for small repulsive $n$ (not shown). The
oscillation increases for large repulsive $n$ as well as for attractive
nonlinearity. From Figs. 1 and 2 we see that for small $n$ the oscillation
is severe for $t>10000$ whereas for large $n (=100)$ it is disturbing even
for $t=100$.  From Figs.  2 we find that a direct numerical solution for
the first excited state of Eq. (\ref{d1}) is the stationary dark soliton
that we look for.

\begin{figure}
 
\begin{center}
\includegraphics[width=.8\linewidth]{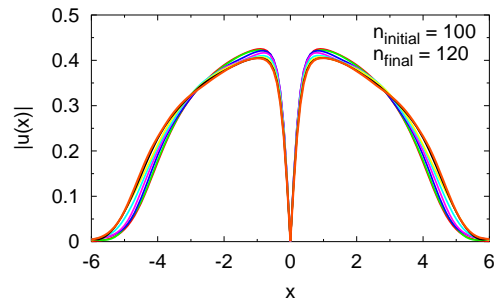}
\end{center}
 
\caption{The   stationary dark
soliton   $|u(x)|$ of Fig.  2 (a)  
at
times $t=0, 10000, 20000, 30000, 40000, 50000$
  calculated
using
the present scheme  with  input (\ref{ex})    
as $n$ is suddenly changed from  100 to 120
at $t=0$.}

\end{figure}

The fact that a certain numerical scheme converges to a
stationary state may not necessarily signify its stability.
Although the 
dark solitons in our study are formed as eigenstates of the
nonlinear equation (\ref{d1}), one needs to demonstrate their stability 
under perturbation.
First we suddenly modify the nonlinearity $n$ of  Eq. (\ref{d1}) after the
dark
soliton is obtained and study the resultant dynamics
for  the dark soliton of  Fig.    
2 (a). In the case of the dark soliton of Fig.  2
(a) calculated 
using our scheme based on  Eq. (\ref{ex}) we suddenly jump the
nonlinearity $n$ from 100 to
120 after the soliton is formed and observe its dynamics for 50000 units
of
time. The resultant dynamics is shown in Fig.   3.

From Fig.  3 we find that even after giving a perturbation by 
changing
$n$, the resultant   stationary dark soliton remains stable for a  large
interval of time (50000 units of time) performing small breathing
oscillation during which
the central notch or minimum of the   stationary dark soliton remains
absolutely stable at $x=0$. From Figs. 2 we find that in the iterative
time evolution method based
on ansatz (\ref{ds2})
the dark
soliton may develop dynamical instability with the central notch
executing
quasi-periodic oscillation around  $x=0$
on time evolution before being destroyed
without any perturbation
whatsoever in a much smaller interval of time than that considered in
Fig.   3.

In addition to the perturbation studied in Fig.  3  we make two
different types of perturbation to strengthen our claim of stability. First 
we  study the dynamics by
increasing  the strength of the harmonic trap by $20\%$: (i)  $x^2 \to 1.2
x^2$. These perturbations are symmetric around $x=0$ which may not
displace the notch in the dark soliton from $x=0$. We consider also the
asymmetric perturbation (ii) $u(x) \to u(x) + 0.02\times \mbox{abs}(u(x))$
where
abs denotes absolute value. As $u(x)$ for the dark soliton is
antisymmetric around $x=0$,  perturbation  (ii) destroys the symmetry
around
$x=0$.     From
Figs.  2 and 3 it is realized that it would be more difficult to have
stability for a large nonlinearity. Hence in the next  two studies on
stability  we
consider
only $n=100$.  The remarkable stability of the soliton under
perturbations (i) and (ii) above is illustrated  in Figs. 4 (a) and
(b). Of
these even the asymmetric perturbation  (ii)   leaves the central notch
of the   stationary dark soliton essentially 
stable at $x=0$. The dynamics at small  times 
in Fig.  4 (b) shows some asymmetry, which, instead of increasing,  
disappears at large times. 
From Fig.  2 (a) we find that the dark soliton calculated
using ansatz (\ref{ds2}) is destroyed after a small time of $t=100$
without
any perturbation whatsoever. 

There seems to be one difference between the behavior of the dark soliton
in Figs. 4 and the corresponding behavior of a BEC in a trap. With a
sudden change of the trap frequency, a BEC executes breathing oscillation
where the central density fluctuates \cite{11a}. In Fig. 4 (a) we also
observe  a similar breathing oscillation. However, in this case because of
the
nature of the black soliton the central density remains zero, at least for
a small change of trap frequency.  In Fig. 4 (b) the dark soliton is given 
a asymetric displacement initially. Because of the stationary nature of
the initial dark soliton, the initial oscillation tends to disappear after
some time as the dark soliton settles to a stationary configuration. The
scenario in Figs. 4 should change upon a large perturbation.

\begin{figure}
 
\begin{center}
\includegraphics[width=.47\linewidth]{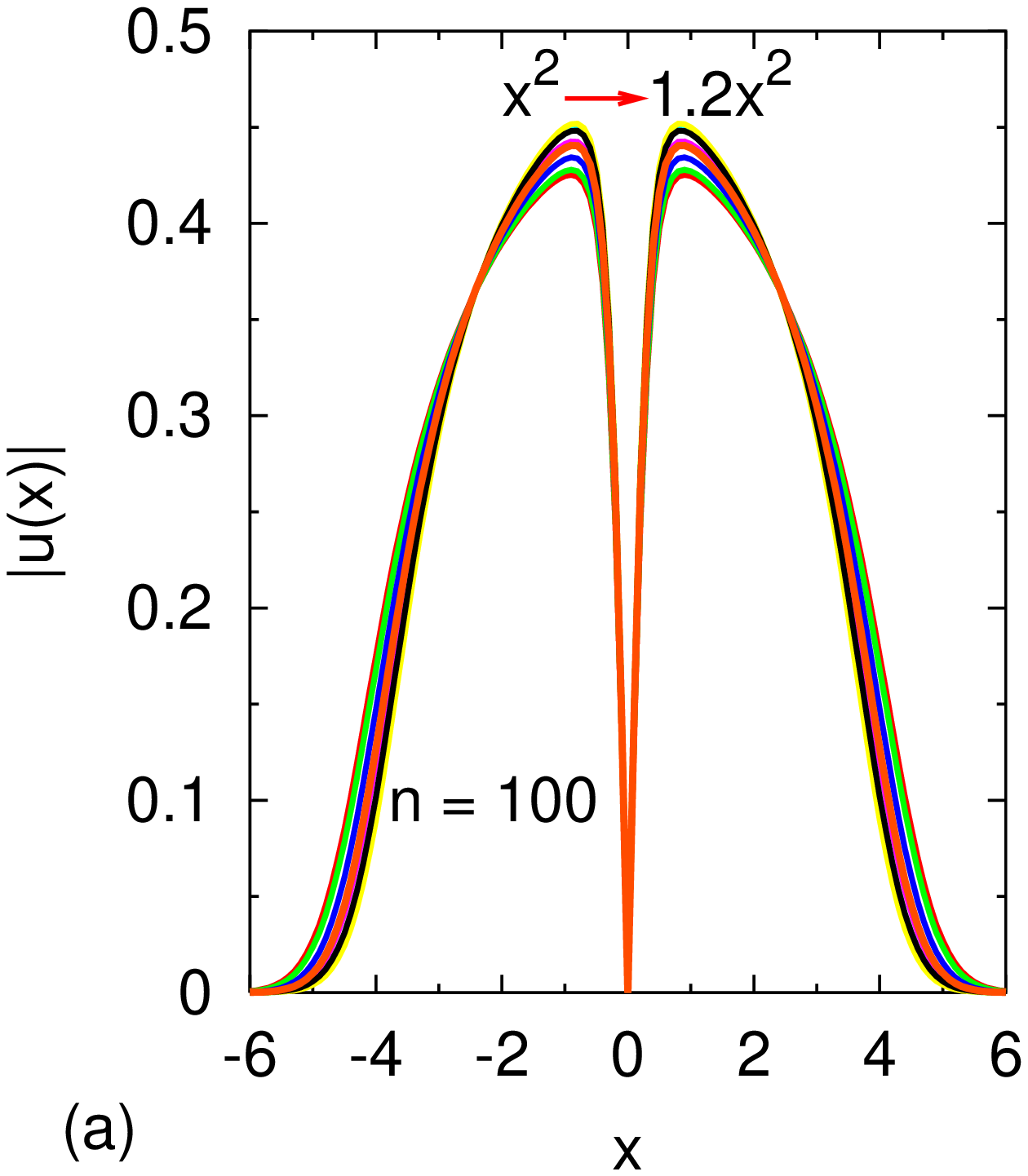}
\includegraphics[width=.47\linewidth]{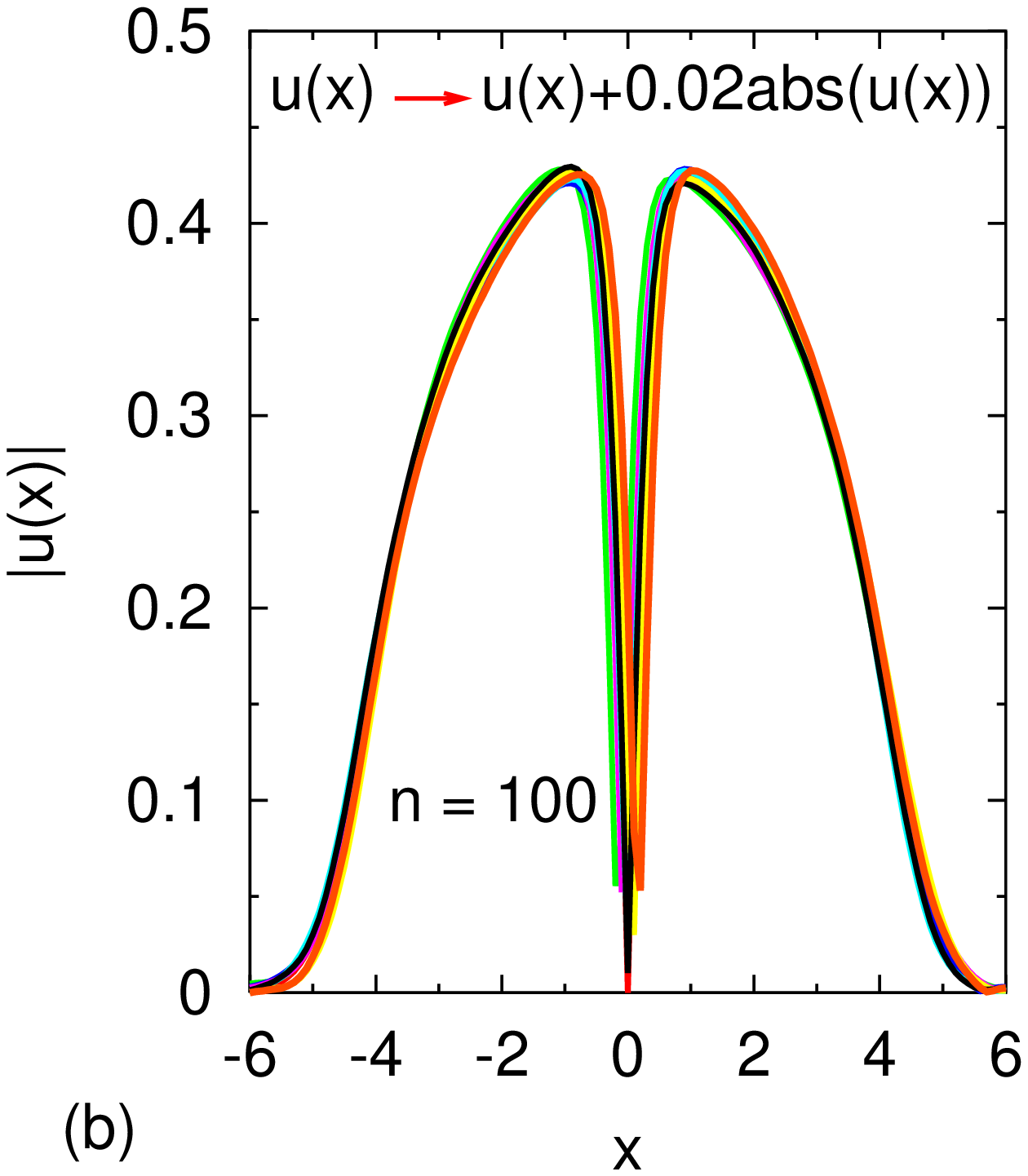}
\end{center}
 
\caption{The dark
soliton   $|u(x)|$ of Fig.   2 (a) at
times $t= 10000, 20000, 30000, 40000, 50000$  calculated
using
the present scheme with  input (\ref{ex})    
(a)
as the harmonic oscillator potential $x^2$ is increased by $20\%: x^2 \to 
1.2 x^2$  and (b) under the change $u(x) \to u(x) + 0.02\times
\mbox{abs}(u(x))$
at
 $t=0$.}

\end{figure}

\section{Harmonic Plus
Optical-Lattice Traps} Now we consider the   stationary dark soliton in a
harmonic
plus
optical-lattice traps. A periodic optical-lattice trap is usually
generated  by
a standing-wave laser beam of wave length $\lambda$. In experiments 
the following superposition of a harmonic plus optical-lattice
traps has been used \cite{9,cata}:
\begin{equation}\label{pot}
V(x)=k x^2+V_0\sin^2(2\pi x/\lambda).
\end{equation} 
Here $V_0$ is
the strength of the optical-lattice potential. We have introduced a parameter $k$ to
control the strength of the harmonic potential.

The search for a   stationary dark soliton in potential 
(\ref{pot}) is performed by introducing this potential in  Eq. (\ref{d1}).
For this purpose we use the input $u_1(x,0)$   of
Eq.  (\ref{ex}) in
Eq.  (\ref{d1}) with $n=V_0=0$
and perform time evolution using the Crank-Nicholson
scheme \cite{11a}. Again the discretization was performed with a space
step 0.05 and time step 0.0025
except for the calculation reported in Fig.   6
where we had to take a space step 0.02 and time step 0.0004. 
 In the course of time evolution the appropriate
nonlinearity and
the optical-lattice potential are switched on  slowly. Then the time
evolution of the 
resultant equation is carried on until a converged solution is obtained.
The results of the calculation for different $n$ are shown in
Figs.   5. These
results  are stationary and do not change with
time evolution in the interval $t=0$ to $t=$ 100000.

However, it was more difficult to obtain  convergence  
when nonlinearity $n$ or strength  $V_0$
increases past 40. For small $n (= 1,10)$ convergence could be
easily 
obtained for $V_0$ up to 80 or so. For larger $n (=20)$ we could obtain
convergence for $V_0$  up to 40 or so. When both $n$ and $V_0$ are
increased,  smaller space and time steps are needed for convergence.
 This is 
understandable as a large nonlinearity 
with a strong    optical-lattice potential 
could seriously jeopardize the numerical accuracy. 
There is no such difficulty 
if the optical-lattice potential 
is removed. If $\lambda$ is reduced there is no convergence 
difficulty, so long as a finer discretization mesh is used.

\begin{figure}

\begin{center}
\includegraphics[width=.47\linewidth]{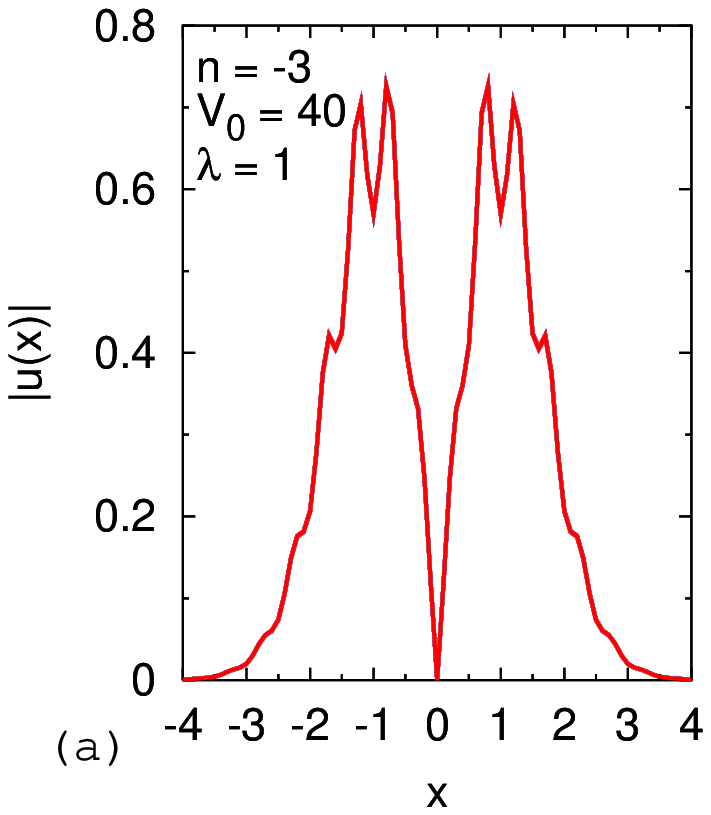}
\includegraphics[width=.47\linewidth]{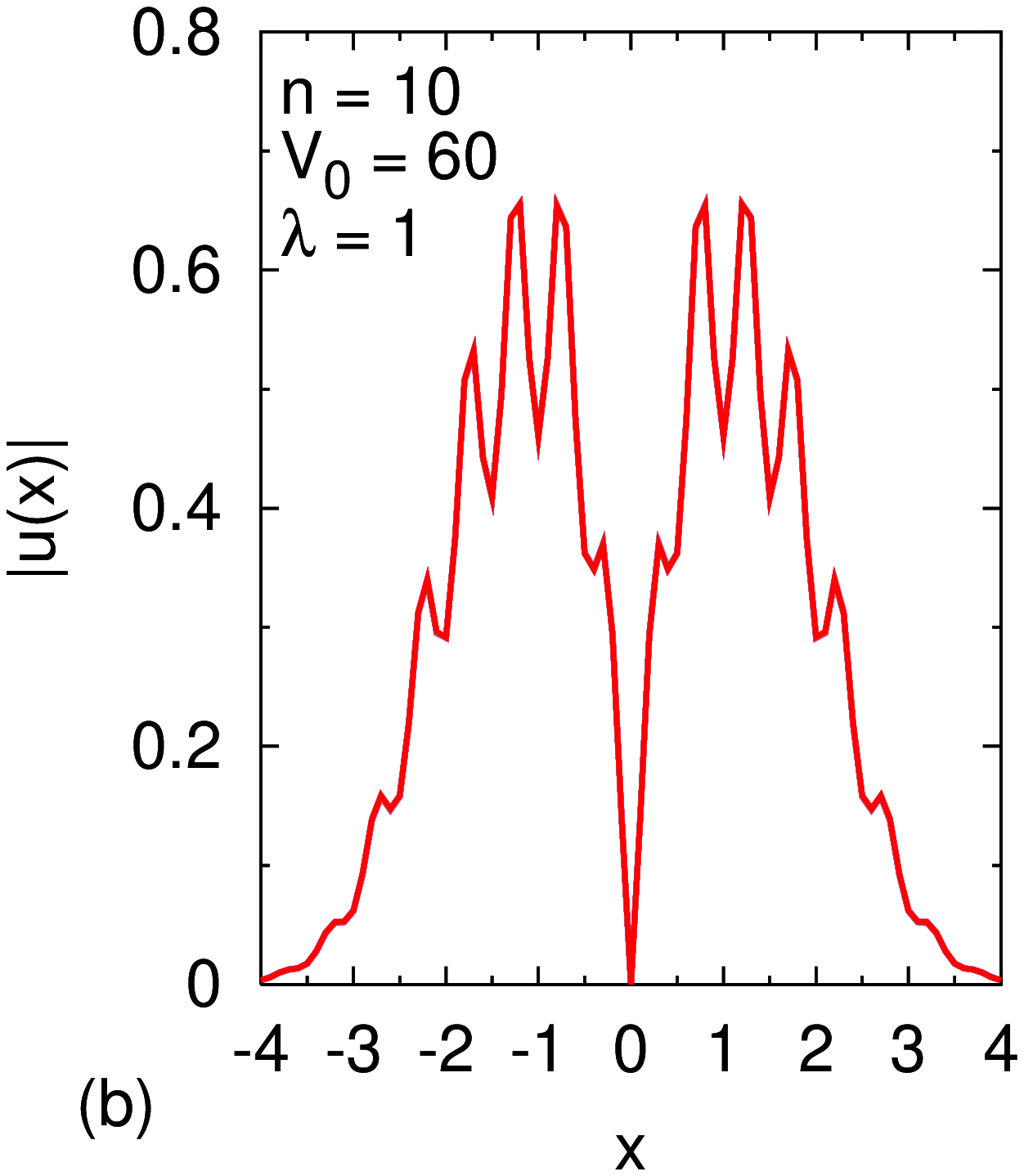}
\end{center}
 
\caption{The   stationary dark soliton   $|u(x)|$   
of Eq.  (\ref{d1}) with potential (\ref{pot}) with $k=\lambda=1$
vs. 
$x$  for $n=$ (a)  $-3$ and  (b) 10  with
strength 
$V_0$ of potential (\ref{pot}) 
given in 
respective figures. The plotted wave function remains 
stationary 
for time evolution of  
Eq.  (\ref{d1})    over 100000 units of time $t$.
}
\end{figure}

The possibility of generating dark solitons in a BEC with a harmonic plus
optical-lattice traps (\ref{pot}) has also been investigated by 
other authors. Kevrekidis { et al.} \cite{9} 
obtained stable solitons for their parameters (chemical
potential, $\lambda$ etc.) for a very weak trap $V(x)=kx^2$ with
$k<0.01$. In this study, we could obtain
such solutions for our parameter values including $k=1$.

\begin{figure}

\begin{center}
\includegraphics[width=0.9\linewidth]{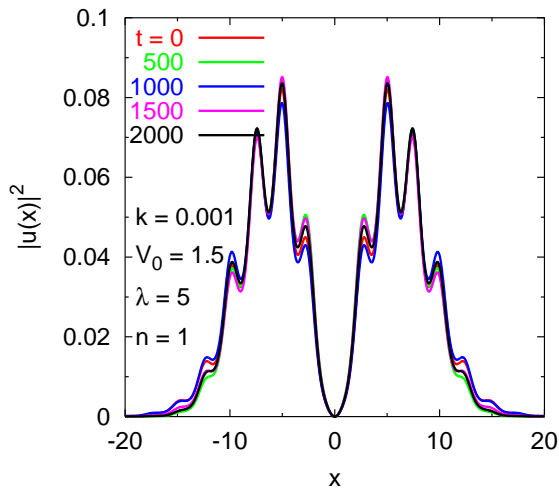}
\end{center}
 
\caption{The probability density   $|u(x)|^2$ 
of  the   stationary dark (black) soliton  of 
Eq.  (\ref{d1}) with potential (\ref{pot}) with $k=0.001$,
 $V_0=1.5$ and $\lambda =5$  vs. $x$ at different times. The dark soliton 
with a zero at $x=0$  
remains stable  for a long time time without executing
quasi-periodic oscillation. 
}
\end{figure}

In view of the present study, it seems that the
instability noted in Ref.  \cite{9} could be due to the use of an initial 
non-stationary state, e. g.  (\ref{ds2}),  to generate the final stationary
dark
soliton.
To substantiate
our claim  we consider a specific case highlighted by Kevrekidis
{ et al.} \cite{9} as an example of instability of the dark soliton in
an
harmonic plus optical-lattice traps, e.g., for $V(x)=kx^2$ with $k=0.001$, 
$V_0=1.5$, and $\lambda =5$ in  Eq. (\ref{pot}).

We repeated the above calculation of Kevrekidis { et al.} \cite{9} 
using our approach. Because of the small value of $k$, the size of the
condensate
is much larger in this case compared to the condensates studied above
and we had to take a much larger number of mesh points  
which results in  large computing time and slower convergence. However,
no quasi-periodic oscillation 
of the position of the minimum of the  
dark soliton  was observed at large times. The present result is
shown in Fig.   6 at different times. 
The central density in our calculation  remains 
strictly zero over large
time scales ($t>2000$), whereas the calculation of Kevrekidis { et al.}
\cite{9} becomes unstable for $t>500$ with the notch of the dark soliton
executing quasi-periodic oscillation. 
There is a difference in the two results, however, which prohibits us to
make a quantitative comparison of the two calculations. In the present
work the 
the wave function is normalized to unity, whereas in Ref. \cite{9} 
the chemical potential has been fixed
to unity. However, here we did not use a random perturbation to test the
stability of the black soliton.

\section{Black Soliton in Three Dimensions} To further fortify  the  claim
of
numerical stability of our scheme we
next test
it in an axially-symmetric
three-dimen\-sional BEC under harmonic as well as optical-lattice traps.
We consider the following  
GP
equation 
for the BEC wave
function $\psi(x,y;t)\equiv \phi(x,y;t)/x$ at radial
position $x$, axial position $y$ and time $t $: \cite{cata2}
\begin{eqnarray}\label{d2}
&\biggr[&- i\frac{\partial
}{\partial t} -{\frac{\partial ^2}{\partial x^2} +\frac{1}{x}
\frac{\partial }{\partial x}- \frac{\partial ^2}{\partial y^2}
  }+\frac{1}{4}
(x^2+\nu^2 y^2) - \frac{1}{x^2}
 \nonumber \\ &+&
 \frac{4\pi^2\kappa}{\lambda^2}\cos^2\frac{2\pi y}{\lambda} +
n\left|\frac
{\phi({ x,y};t)}{x}\right|^2
 \biggr]\phi({  x,y};t)=0,
\end{eqnarray}with normalization
\begin{equation}
2\pi \int_{-\infty}^{\infty}dy \int_0^\infty x dx
|\psi(x,y;t)|^2=1,
\end{equation}where $(x^2+\nu^2 y^2)/4 $ is the axial  harmonic trap
and $(4\pi^2\kappa/\lambda^2)$ $
\cos^2(2\pi y/\lambda) $
the optical-lattice potential. Here
length and time are expressed in units of  $l/\sqrt 2 \equiv \sqrt{
\hbar/(2m\omega)}$ and $\omega
^{-1}$. respectively, with $\omega $ the radial trap frequency, $m$
the atomic mass, $\nu$ the axial parameter,
and
$n=8 \pi \sqrt 2 N_0a/l$  the scaled nonlinearity, where $N_0$ is the
number
of atoms and  $a$ the scattering length. The numerical solution is
calculated by the Crank-Nicholson discretization scheme where
we used as in Ref.  \cite{11a} a space step of 0.1 in both radial and
axial directions and a time step of 0.001.
The   stationary dark soliton we are looking for is a nonlinear extension
of
the following solution of  Eq. (\ref{d2}) with $n=\kappa=0$:
\begin{equation}\label{tr}
\phi(x,y)=\left( \frac{\nu}{2\pi}\right)^{3/4}
x y  \exp[-(x^2+\nu y^2)/4],
\end{equation}
and can be found in a  regular fashion by time evolution of
Eq. (\ref{d2}) with $n=\kappa=0$ with  Eq.  (\ref{tr}) as the initial
solution. This solution has a notch at $y=0$ in the axial
$y$ direction. In the course of time evolution the nonlinearity $n$ and
the optical-lattice strength $\kappa$ are slowly introduced until the
final values of these parameters are attained.

\begin{figure}
 
\begin{center}
\includegraphics[width=\linewidth]{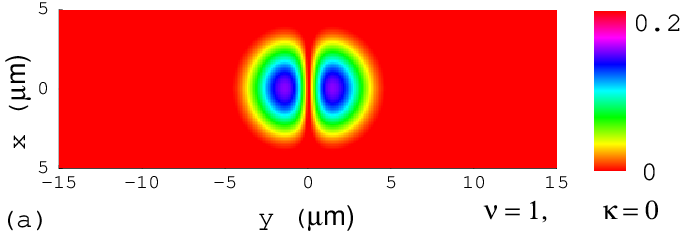}
\includegraphics[width=\linewidth]{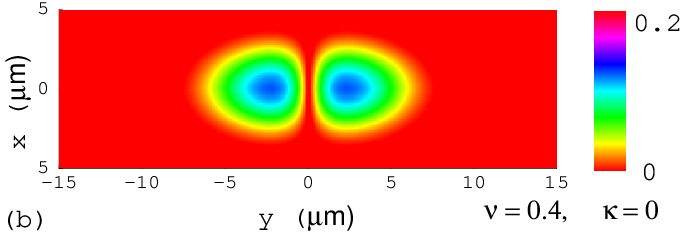}
\includegraphics[width=\linewidth]{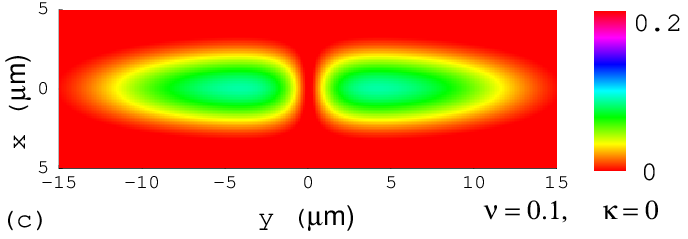}
\end{center}
 
\caption{Contour plot of
the wave function   $|\psi(x,y)|$ of the  black soliton of
Eq.  (\ref{d2}) with harmonic trap alone for 
(a) $\nu =1$, (b)   $\nu =0.4$, and (c)  $\nu =0.1$, and 
$n=100$.  }
\end{figure}

\begin{figure}
 
\begin{center}
\includegraphics[width=\linewidth]{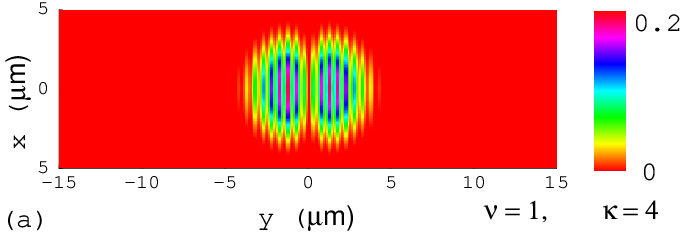}
\includegraphics[width=\linewidth]{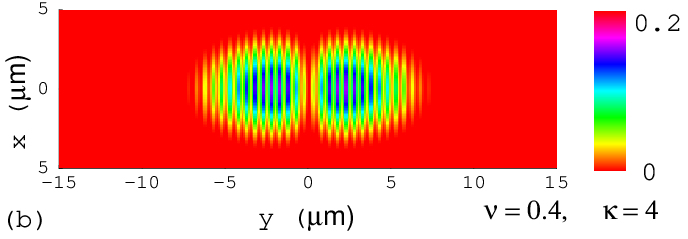}
\includegraphics[width=\linewidth]{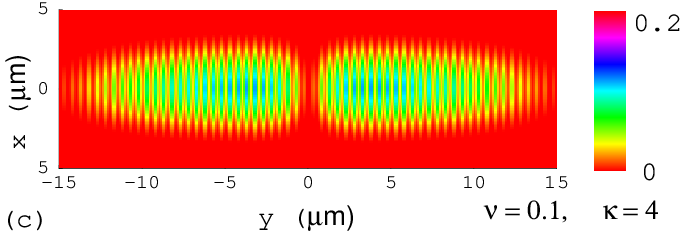}
\end{center}
 
\caption{Contour plot of
the black-soliton wave function   $|\psi(x,y)|$ of
Eq.  (\ref{d2}) with  with $\kappa=4$,
$\lambda =1$ and 
(a) $\nu =1$, (b)   $\nu =0.4$, and (c)  $\nu =0.1$, and  
$n=100$.  } 
\end{figure}

 In this study we take the atoms to be $^{87}$Rb and a final nonlinearity
$n=
100$ together with $\omega = 2\pi \times 90$ Hz so that $l/\sqrt 2 \approx
0.8$ $\mu$m. First we study the generation of a   stationary dark soliton
in the
harmonic trap alone by setting $\kappa =0 $ in Eq. (\ref{d2}) for three
values
of $\nu=1,0.4$ and 0.1.  Contour plots of the wave function $\psi(x,y)$
of the   stationary dark
soliton in these three cases  are exhibited in Figs.  7 (a), (b), and (c),
respectively, where 
the central notch appears
prominently and does not oscillate upon time evaluation. 
 
Next we study the   stationary dark soliton in the above problem in the
presence of an
optical-lattice potential with $\kappa=4$ and $\lambda =1$ in addition to
the above harmonic potential. The contour plots of the   stationary dark
soliton in
this case are shown in Figs. 8.  In addition to the more prominent notch
at the center signaling a   stationary dark soliton, there are periodic
lines along the
axial direction due to the optical-lattice potential. This should be
contrasted with the similar wave function in one dimension exhibited in
Figs. 5 and 6, where there is also periodic modulation of the wave
function due
to the optical-lattice potential. Needless to say that these dark solitons
in three-dimensions are stationary solutions of the GP equation and do not
exhibit instability upon time evolution.

\section{Conclusion}

We show that 
the   stationary dark (black) soliton of a trapped zero-temperature BEC is
actually a
nonlinear
extension of the first vibrational excitation 
of the  linear
problem 
obtained by setting $n=0$ in Eq. (\ref{d1}). Based on this, we
suggest a time-evolution calculational scheme starting from the 
linear problem with the harmonic potential alone
while the nonlinearity and any additional 
potential
(such as the optical-lattice potential)
are slowly introduced during time
evolution. This results in a stable numerical scheme for the
stationary dark
soliton as during time evolution the system always passes through
successive stationary 
eigenstates of the nonlinear GP equation until the desired
stationary dark soliton
is
obtained also as a  stationary     eigenstate of the final GP equation.
The present
approach is equally applicable to  both 
repulsive and attractive interactions in one and three dimensions and 
eliminates the so called dynamical instability of the
dark
soliton in the presence of  a trap 
and yields a robust   stationary dark soliton. 
To demonstrate the robustness of the present numerical scheme, we
illustrate the
stable prolonged breathing oscillation of the dark soliton upon the
application of a perturbation. We have presented our results in a
dimensionless time unit. In a typical experimental situation this unit
corresponds to about 1 ms. 
The stability of the dark soliton during few thousand units
of time as considered in this paper is sufficient for this
purpose. Time scales beyond a few seconds are completely
unrealistic in this context, since the condensate only lives for about 15
seconds, with the soliton predicted to have a much shorter 'lifespan' due
to
thermal and quantum fluctuations even at very low temperatures.

\vskip .5cm

 
\noindent{We thank   Dr. V. V. Konotop and  
Dr. N. P. Proukakis
    for helpful e-mails. 
The work was supported in part by the CNPq and FAPESP 
of Brazil.}

\end{document}